\input{aipcheck}

\documentclass[
    ,final            
  ]
  {aipproc}

\layoutstyle{8x11double}

\usepackage{bm}

\def\A{{\cal{N}}}

\begin{document}

\title{Conservation of nonlinear curvature perturbations on super-Hubble scales}


\author{Misao Sasaki}{
  address={Yukawa Institute for Theoretical Physics, Kyoto University,
Kyoto 606-8502, Japan}
}

\begin{abstract}
We consider general, non-linear curvature perturbations on scales 
greater than the Hubble horizon scale by invoking
an expansion in spatial gradients, the so-called gradient expansion.
After reviewing the basic properties of the gradient expansion,
we derive the conservation law for non-linear curvature perturbations
for an isentropic fluid. We also define the gauge-invariant curvature
perturbation under a finite shift of time-slicing, and derive the
non-linear genralization of the $\delta N$ formalism.
The results obtained are straight-forward
generalisations of those already proven in linear perturbation theory,
and the equations are
simple, resembling closely the first-order equations.
\end{abstract}

\maketitle


\section{Introduction}

Analysis of the WMAP first-year data~\cite{Bennett:2003bz} revealed that
the standard inflationary universe scenario in which a single
inflaton field slowly rolls down its potential hill during
inflation is perfectly consistent with the observed CMB
temperature spectrum~\cite{Hinshaw:2003ex}.
Namely, the CMB anisotropy theoretically
predicted from a scale-invariant adiabatic perturbation
in a spatiall flat universe matched with the observed CMB
anisotropy with impressive precision~\cite{Spergel:2003cb}, and no non-Gaussianity
was found in the statistics of the observed anisotropy~\cite{Komatsu:2003fd}.
This not only means the validity of the standard inflationary
theory but also justifies the use of linear cosmological 
perturbation theory~\cite{Bardeen80,KodSas,Mukhanov}.

Nevertheless, there are some features in the observed
CMB spectrum which might not be caused simply
by cosmic variance but that could be a signature of non-standard inflation such as
multi-field inflation, non-slow-roll inflation,
braneworld inflation, etc.. Also, the level of 
the Gaussianity test is still not stringent enough
that there could be some non-Gaussianity which are
to be found in future experiments, say, by PLANCK~\cite{Planck}.
To deal with such cases properly, it is necessary to 
consider nonlinear perturbations on superhorizon scales.

Turning to the state of our present universe, the WMAP data also confirmed
the existence of energy in the form of a cosmological constant or
vacuum energy, now called the dark energy, and it dominates the energy
density of the present universe. Because the energy scale of this vacuum
(dark) energy is extremely small, of $O(10^{-3})$eV,
compared to the Planck scale, $10^{19}$Gev, the confirmation of
its existence was a big shock to theorists, especially
those from the particle physics/string theory community.

Recently, several authors claimed that the present dark energy
can be explained by backreaction of super-Hubble scale perturbations
at second order~\cite{Barausse:2005nf, Kolb:2005me}. 

Nonlinear dynamics of cosmological perturbations on superhorizon
scales has been discussed already by many authors~\cite{gradient}. 
But in the light of these new developments mentioned above, it is
appropriate to revist the dynamics of cosmological perturbations on
superhorizon scales, and clarify what can be said and what cannot be said.

In this report, we give a fully nonlinear formulation of
cosmological perturbations on superhorizon scales at
leading order in the gradient expansion, based on
our recent paper~\cite{LMS}.

\section{Gradient expansion}

The gradient expansion is a powerful tool
to deal with cosmological perturbations on superhorizon scales.
It was developed by various authors for various purposes~\cite{gradient}.
Here we first briefly describe the essence of it.

The gradient expansion assumes that spatial derivatives are
always smaller than the time derivative for any physical quantity
$Q$:
\begin{eqnarray}
\left|\frac{\partial Q}{\partial x^i}\right|
\ll \left|\frac{\partial Q}{\partial t}\right|.
\end{eqnarray}
In cosmological situations, we have $|\partial_tQ|\sim HQ$
where $H$ is the inverse of the Hubble timescale
(or the gravitational free-fall timescale) and given by
$H\sim \sqrt{8\pi G\rho/3}$. At each point of spacetime,
we therefore have a local definition of the Hubble horizon 
size, given by $H^{-1}$. Hence the gradient
expansion is valid if the quantities are slowly varying
in space on the Hubble horizon scale.

\begin{figure}
  \includegraphics[width=0.8\textwidth]{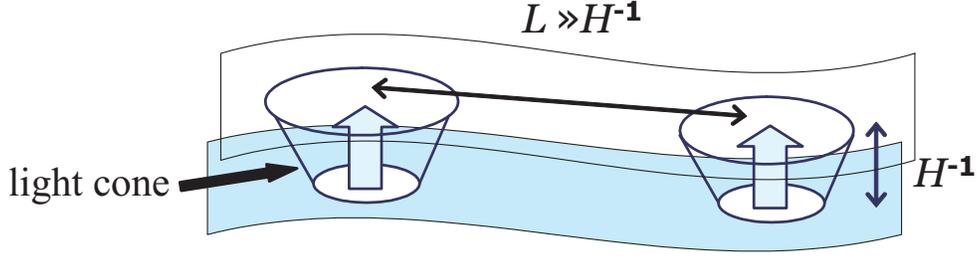}
  \caption{Schematic spacetime diagram for superhorizon scale
   cosmological perturbations. Each `local' Hubble-horizon region 
   evolves independently because of causality.}
  \label{lightcone}
\end{figure}

At lowest order in the gradient expansion, the 
Einstein and matter field equations become ordinary
differential equations in time. In other words,
the evolution on scale of each Hubble size region,
which we call `local' evolution, should not depend
on what is happening in some spatially distant part
of the universe, as illustrated in Fig.~\ref{lightcone}.
 This is just a consequence of causality.

\subsection{Metric}

We consider the standard $(3+1)$-decomposition of the metric, 
which applies to any smooth spacetime:
\begin{eqnarray}
ds^2 = -\A^2dt^2+ \gamma_{ij}(dx^i+\beta^idt)(dx^j+\beta^jdt)
\,,
\label{metric0}
\end{eqnarray}
where $\A$ is the lapse function, $\beta^i$ the shift vector, and
$\gamma_{ij}$ the spatial three metric. 
(Greek indices will take the values $\mu,\nu=0,1,2,3$, Latin indices
$i,j=1,2,3$. The spatial indices are to be raised or lowered by $\gamma^{ij}$
or $\gamma_{ij}$.) In this $(3+1)$-decomposition, the unit time-like vector
normal to the $x^0=t=$const. hypersurface $n^\mu$ has the components,
\begin{eqnarray}
n_\mu=\left[-\A,0\right],
\quad n^\mu=\left[\frac{1}{\A},-\frac{\beta^i}{\A}\right].
\label{normal}
\end{eqnarray}
We write the 3-metric, $\gamma_{ij}$, as a product of two terms,
\begin{eqnarray}
\label{defgammaij}
\gamma_{ij}\equiv e^{2\alpha}\tilde\gamma_{ij} \,,
\end{eqnarray}
where $\alpha$ and $\tilde\gamma_{ij}$ are functions of
the spacetime coordinates $(t,x^i)$, and $\det[\tilde\gamma_{ij}]=1$. 

The above form of the metric is completely general.
Now we apply it to the case of our interest, i.e., to
cosmological situations in which the metric varies substantially
on scales much greater than the Hubble horizon but is almost
homogeneous and isotropic (and spatially flat) on scales of the
Hubble horizon. Then, although it is not necessary, it is convenient
to introduce a fiducial `background' scale factor $a(t)$ and
a perturbation $\psi$ on it as
\begin{eqnarray} 
e^\alpha = a(t) e^{\psi(t,x^i) }\,.
\label{alpha}
\end{eqnarray}
Note that the variable $\psi$ describes the curvature perturbation
in the limit of linear theory,
\begin{eqnarray}
{}^{(3)}R(\gamma)=-\frac{4}{a^2}{}^{(3)}\Delta\,\psi+O(\psi^2)\,.
\end{eqnarray}
Hence we also call it the curvature perturbation, although there
will be no simple relation between $\psi$ and ${}^{(3)}R$ 
in the nonlinear case.

Likewise, it is convenient to factorize the matrix
$\tilde\gamma_{ij}$ as
\begin{eqnarray}
\tilde\gamma  \equiv I e^H
\,,
\end{eqnarray}
where $I$ is the unit matrix.
The condition $\det(\tilde\gamma)=1$  ensures that the matrix
$H$ is traceless. In the limit of linear theory,
the transverse part of $H_{ij}$ describes the tensor
(gravitational wave) perturbation. 

To invoke the gradient expansion, we associate an
expansion parameter $\epsilon$ with each spatial derivative,
$\partial_i\to \epsilon\partial_i$. Physically, $\epsilon$
is equal to the ratio of the Hubble horizon size to the
typical wavelength of a perturbation, $\epsilon\sim H^{-1}/\lambda$. 
As for the form of the metric (\ref{metric0}) is concerned,
in order to apply the gradient expansion, the only non-trivial
assumption we make is that
\begin{eqnarray}
\beta^i=O(\epsilon)\,.
\label{beta}
\end{eqnarray}

\subsection{Matter}

The energy-momentum tensor is assume to have the perfect fluid form
\begin{eqnarray}
\label{defTmunu}
T_{\mu\nu}\equiv \left(\rho+P\right)u_\mu u_\nu
+g_{\mu\nu}P \,,
\end{eqnarray}
where $\rho=\rho(x^\mu)$ is the energy density and $P=P(x^\mu)$ is the
pressure. 
 Then the energy conservation equation is
\begin{eqnarray}
-u_\mu \nabla_\nu T^{\mu\nu}
=\frac{d}{d\tau}\rho+\left(\rho+P\right)\theta=0\,,
\label{econs}
\end{eqnarray}
where $d/d\tau=u^\mu\nabla_\mu$ and
\begin{eqnarray}
\label{exp}
\theta\equiv\nabla_\mu u^\mu\,.
\end{eqnarray}

In accordance with the assumption (\ref{beta}) on the shift vector
$\beta^i$, we also assume that the 3-velocity of the matter to
satisfy
\begin{eqnarray}
v^i\equiv\frac{u^i}{u^0}=O(\epsilon)\,.
\label{3velocity}
\end{eqnarray}
Then it follows that
\begin{eqnarray}
\theta=\frac{3\partial_t\alpha}{\A}+O(\epsilon^2)=\theta_n+O(\epsilon^2)\,,
\label{theta}
\end{eqnarray}
where
\begin{eqnarray}
\theta_n\equiv \nabla_\mu n^\mu\,.
\end{eqnarray}
That is, the expansion of the matter 4-velocity and that of
the unit vector normal to the hypersurface coincide to
each other at lowest order in the gradient expansion,
which follows from the assumptions~(\ref{beta}) and (\ref{3velocity}).

One can then introduce the notion of the local Hubble parameter 
$\tilde H$ by
\begin{eqnarray}
\tilde H\equiv\frac{1}{3}\theta
=\frac{1}{3}\theta_n+O(\epsilon^2)
=\partial_t\alpha+O(\epsilon^2)\,.
\label{localH}
\end{eqnarray}
The second (or the last) equality implies that this defintion 
is unique
in the sense that the local Hubble parameter can be defined
in terms of the expansion of the hypersurface normal $n^\mu$,
and the definition is independent of the choice of the time slicing.
Here, we note again that `local' means `on scales of the
Hubble horizon size.'

\subsection{Local Friedman equation}

So far, we have not specified the theory of gravity.
Let us now invoke Einstein gravity.
By explicitly writing down the Einstein equations,
one sees that one can further consistently assume~\cite{LMS}
\begin{eqnarray}
\frac{\partial}{\partial t}\tilde\gamma_{ij}=O(\epsilon^2)\,.
\label{tgamma}
\end{eqnarray}
Physically this assumption means the absence of a decaying
mode in the shear of the hypersurface. In the inflationary
universe, this decaying mode dies out rapidly soon after
the comoving scale of interest leaves out of the horizon.
We then find that the only non-trivial equation is the
$(0,0)$-component of the Einstein equations, i.e., the
Hamiltonian constraint on the $t=$constant hypersurface,
which reads
\begin{eqnarray}
3\tilde H^2=8\pi G\rho +O(\epsilon^2)\,.
\label{localF}
\end{eqnarray}
This is exaclty the same as the usual Friedmann equation
for a homogeneous and isotropic, spatially flat universe.
Together with the energy conservation (\ref{econs}), we now
see that the Friedmann equation is locally valid in each 
Hubble-horizon size region even under the presence of
nonlinear perturbations on superhorizon scales.

An immediate consequence of Eq.~(\ref{localF}) is that
the slicing for which the energy density is uniform
on each time slice (uniform density slicing) is 
equivalent to the slicing for which the local Hubble
parameter is uniform (uniform Hubble slicing),
up to errors of $O(\epsilon^2)$. Although not
apparent from Eq.~(\ref{localF}), one can also
easily show from the momentum constraint equations
that the comoving slicing for which 
$T^0{}_j=(u^0)^2(\rho+P)(v_i+\beta_i)=0$ 
coincides also with the uniform Hubble or uniform
density slicing up to errors of $O(\epsilon^2)$~\cite{LMS}.

Another, most important consequence of Eq.~(\ref{localF})
is that the local physics cannot be affected by
the presence of superhorizon scale perturbations,
no matter how large they are. In particular, this 
implies that there will be no modification or backreaction
to the local Friedmann equation due to 
superhorizon scale perturbations, contrary to the claim
made in \cite{Barausse:2005nf, Kolb:2005me}. 
More explicit investigations of the issue have been done
recently by Hwang and Noh~\cite{Hwang} and 
Hirata and Seljak~\cite{Hirata:2005ei}, whose results
are in agreement with our conclusion.

\section{Nonlinear curvature perturbation and $\bm{\Delta N}$ formla}

Now we investigate the evolution of the curvature perturbation
$\psi$ at lowest order in the gradient expansion.
We first show the conservation of nonlinear curvature perturbation
for an isentropic (adiabatic) fluid. We then derive the
$\Delta N$ formula, which relates the amplitude of curvature
perturbation to the perturbation in the $e$-folding number
of cosmic expansion, for the general case.

\subsection{Nonlinear conservation of $\bm\zeta$}

Multiplying the energy conservation
equation (\ref{econs}) by $\A$, we have
\begin{eqnarray}
\label{conti}
\frac{\dot a}{a} + \partial_t\psi 
= -\frac13 \frac{\partial_t\rho}{\rho + P} + O(\epsilon^2)
\,.
\end{eqnarray}
So far, we have not specified the choice of the time-slicing.
Let us now choose the uniform density slicing,
$\rho=\rho(t)$. Following the standard notation~\cite{Wands:2000dp},
we introduce
\begin{eqnarray}
\zeta\equiv -\psi\quad\mbox{on uniform density slices}.
\end{eqnarray}
Then, provided that $P$ is a unique function of $\rho$ 
(the `adiabatic pressure' condition), Eq.~(\ref{conti}) shows that
$\partial_t\psi$ is spatially homogeneous to first order,
\begin{eqnarray}
\label{dotz}
-\partial_t\psi=\partial_t\zeta = O(\epsilon^2)\,,
\end{eqnarray}
where the part of $\psi$ that may be a function of
only time is assumed to be absorbed in the fiducial
background scale factor $a(t)$ without loss of generality.
Thus $\zeta$ is conserved on superhorizon scales.
This is a generalization of conserved curvature
perturbation on uniform density slices in
linear~\cite{Wands:2000dp} and
second order~\cite{LW} to nonlinear order.

It may be noted that, as in linear or second order theory,
the conservation law holds for each fluid for each corresponding
uniform density slicing, if there exist multiple fluids, as 
long as they are interacting only gravitationally, and 
it is derived solefy from the energy conservation law,
independently of the gravitational theory one considers.

\subsection{$\bm{\Delta N}$ formula}

Let us define the number of $e$-foldings of expansion 
along an integral curve of the 4-velocity (a comoving worldline):
\begin{eqnarray}
N(t_2,t_1;x^i)\equiv\frac{1}{3}\int_{t_1}^{t_2}\theta\,\A dt
=-\frac{1}{3}\int_{t_1}^{t_2}dt
\left.\frac{\partial_t\rho}{\rho+P}\right|_{x^i}\,,
\label{Ndef}
\end{eqnarray}
where, for definiteness, we have chosen the spatial coordinates
$\{x^i\}$ to be comoving with the fluid. 
It is very important to note that this definition is purely geometrical,
independent of the gravitational theory one has in mind, and applies
to any choice of time-slicing.

{}From (\ref{localH}) we find
\begin{eqnarray}
\psi(t_2,x^i)-\psi(t_1,x^i)
=N(t_2,t_1;x^i)-\ln\left[\frac{a(t_2)}{a(t_1)}\right]\,.
\label{evolve}
\end{eqnarray}
Thus we have the very general result that the change in $\psi$, going
from one slice to another, is equal to the difference between the
actual number of $e$-foldings and the background value
$N_0(t_2,t_1)\equiv\ln[a(t_2)/a(t_1)]$.  One immediate consequence of
this is that the number of $e$-foldings between two time slices will
be equal to the background value, if we choose the `flat slicing' on
which $\psi=0$.

Consider now two different time-slicings, say slicings $A$ and $B$,
which coincide at $t=t_1$ for a given spatial point $x^i$ of our
interest (i.e., the 3-surfaces $\Sigma_A(t_1)$ and $\Sigma_B(t_1)$ are
tangent to each other at $x^i$). Then the difference in the
time-slicing at some other time $t=t_2$ can be described by the
difference in the number of $e$-foldings. From Eq.~(\ref{evolve}), we
have
\begin{eqnarray}
\psi_A(t_2,x^i)-\psi_B(t_2,x^i)
&=&N_A(t_2,t_1;x^i)-N_B(t_2,t_1;x^i)
\nonumber\\
&\equiv&\Delta N_{AB}(t_2,x^i)\,,
\label{gaugetran}
\end{eqnarray}
where the indices $A$ and $B$ denote the slices $A$ and $B$,
respectively, on which the quantities are to be evaluated.  

\begin{figure}
  \includegraphics[width=0.8\textwidth]{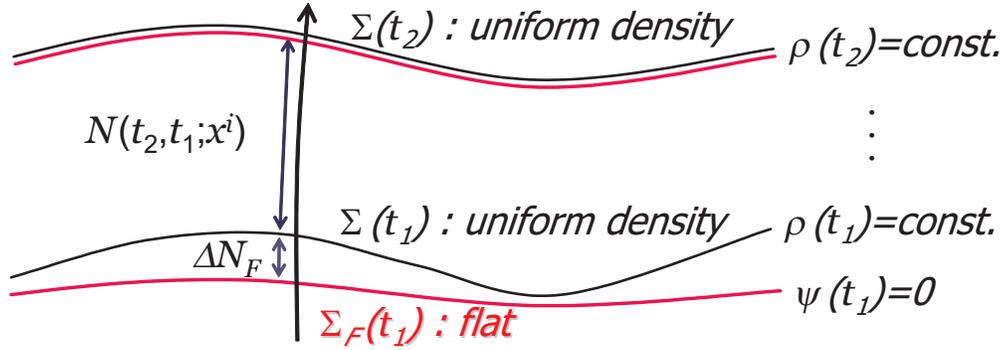}
  \caption{A schematic space-time diagram for the $\Delta N$ formula.
   The slicing that interpolates between $\Sigma(t_1)$ and $\Sigma(t_2)$
   is the uniform density slicing, while the relevant slicing for
   the $\Delta N$ formula is the one that starts from the flat slice
   $\Sigma_F(t_1)$ and ends on the uniform density slice $\Sigma(t_2)$.
   The curvature perturbation on the final hypersurface $\Sigma(t_2)$ is
   given by the difference in the $e$-folding number
   from $\Sigma_F(t_1)$ to $\Sigma(t_2)$ and that from $\Sigma(t_1)$ to
   $\Sigma(t_2)$.}
  \label{deltaN}
\end{figure}

Now let us choose the slicing $A$ to be such that it starts on a flat
slice at $t=t_1$ and ends on a uniform-density slice at $t=t_2$, and
take $B$ to be the flat slicing all the time from $t=t_1$ to $t=t_2$.
Then applying Eq.~(\ref{gaugetran}) to this case, we have
\begin{eqnarray}
\psi_A(t_2,x^i)
=N_A(t_2,t_1;x^i)-N_0(t_2,t_1)=\Delta N_F(t_2,t_1;x^i)\,,
\end{eqnarray}
where $\Delta N_F(t_2,t_1;x^i)$ is the difference in the number of
$e$-foldings (from $t=t_1$ to $t=t_2$) between the uniform-density
slicing and the flat slicing, as illustrated in Fig.~\ref{deltaN}.
  This is a non-linear version of the
$\Delta N$ formula that generalises the first-order result of Sasaki
and Stewart

Now we specialise to the case $P=P(\rho)$. In this case,
Eq.~(\ref{evolve}) reduces to
\begin{eqnarray}
\psi(t_2,x^i)-\psi(t_1,x^i)
=-\ln\left[\frac{a(t_2)}{a(t_1)}\right]
-\frac{1}{3}\int^{\rho(t_2,x^i)}_{\rho(t_1,x^i)}\frac{d\rho}{\rho+P}\,.
\end{eqnarray}
Thus, there is a conserved quantity, which is independent of the
choice of time-slicing, given by
\begin{eqnarray}
-\zeta(x^i)\equiv\psi(t,x^i)
+\frac{1}{3}\int_{\rho(t)}^{\rho(t,x^i)}\frac{d\rho}{\rho+P}\,.
\label{nlconserve}
\end{eqnarray}
In the limit of linear theory, this reduces to the conserved
curvature perturbation in the uniform-density, uniform-Hubble, or 
the comoving slicing,
\begin{eqnarray}
-\zeta(x^i)={\cal R}_c(x^i)
=\psi(t,x^i)+\frac{\delta\rho(t,x^i)}{3(\rho+P)}\,,
\label{linear}
\end{eqnarray}
where ${\cal R}_c$ is the curvature perturbation on 
the comoving slices~\cite{SaSt95}.

\section{conclusion}

We have investigated the behavior of nonlinear cosmological
perturbations on superhorizon scales by invoking
the gradient expansion. Already at lowest order in the
expansion, we have obtained some non-trivial, very useful results.
They are summarized as follows:
\begin{list}{$\bullet$}{}
\item 
The Friedmann equation for a spatially flat universe holds
locally (on scales of the Hubble horizon size), no matter
how big the perturbation is on superhorizon scales. 
A direct consequence of this is that there will be no backreaction effect 
from super-Hubble perturbations on local (horizon-size) physics.
Local physics is determined solely by local physics.

\item
There exists a non-linear generalization of $\zeta$, which describes
the curvature perturbation on uniform density slices, which is conserved 
for a barotropic fluid on super-Hubble scales.

\item 
There exists a non-linear generalization of $\Delta N$ formula,
which relates the final amplitude of the curvature perturbation
to the difference in the $e$-folding number between 'flat'
and uniform density slices at an initial epoch (which may be
chosen arbitrarily provided that the comoving scale of interest
is beyond the Hubble-horizon scale).
This formula may be useful in evaluating non-Gaussianity from inflation.

\end{list}

In this report, we only discussed the properties of nonlinear
perturbations on superhorizon scales at lowest order in the gradient
expansion. However, in non-standard models of inflation, such as in
a non-slow-roll model, the second order corrections in the gradient
expansion is known to be important already in linear perturbation
theory~\cite{Starobinsky:1992ts,Leach:2001zf}. Thus extending
the present analysis to second order in the gradient expansion
will be necessary to deal with more general cases.
This issue is under investigation~\cite{YTanaka}.


\begin{theacknowledgments}
I would like to thank the organizors of PASCOS05 for inviting
me to present this work at the meeting. This work is based on
collaboration with David Lyth and Karim Malik~\cite{LMS}.
This work was supported
in part by Monbukagakusho Grant-in-Aid for Scientific 
Research(S) No. 14102004 and (B) No.~17340075.
\end{theacknowledgments}






\end{document}